\documentclass[cameraready]{Interspeech}
\usepackage{comment}

\title{LLM-Guided Reinforcement Learning for Audio-Visual Speech Enhancement}

\author[affiliation={1}]{Chih-Ning}{Chen}
\author[affiliation={2}]{Jen-Cheng}{Hou}
\author[affiliation={2}]{Hsin-Min}{Wang}
\author[affiliation={1}]{Shao-Yi}{Chien}
\author[affiliation={2}, correspondingauthor]{Yu}{Tsao}
\author[affiliation={3}] {Fan-Gang}{Zeng}

\address{
    $^1$ Department of Electrical Engineering, National Taiwan University, Taiwan \\
    $^2$ Research Center for Information Technology Innovation, Academia Sinica, Taiwan\\
    $^3$ Center for Hearing Research, University of California Irvine, USA
}

\email{\{ning, sychien\}@media.ntu.ee.edu.tw, \{jchou,  whm, yu.tsao\}@citi.sinica.edu.tw, fzeng@uci.edu}


\begin{document}
\maketitle

\begin{abstract}
In existing Audio-Visual Speech Enhancement (AVSE) methods, objectives such as Scale-Invariant Signal-to-Noise Ratio (SI-SNR) and Mean Squared Error (MSE) are widely used; however, their correlation with perceived speech quality is often suboptimal and provides limited interpretability for optimization. This work proposes a reinforcement learning–based AVSE framework with a Large Language Model (LLM)-based interpretable reward model. An audio LLM generates natural language descriptions of enhanced speech, which are converted by a sentiment analysis model into a 1–5 rating score serving as the PPO reward for fine-tuning a pretrained AVSE model. Compared with scalar metrics, LLM-generated feedback is semantically rich and explicitly describes speech quality improvements. Experiments on the AVSEC-4 dataset show that the proposed method outperforms a supervised baseline and a DNSMOS-based RL baseline in PESQ, STOI, neural quality metrics, and subjective listening tests.

\end{abstract}

\section{Introduction}

In real-world environments, speech is often corrupted by background noise. Speech Enhancement (SE) techniques are therefore widely used to suppress noise and improve speech quality and intelligibility \cite{lu2013speech, wang2018supervised, o2024speech}. Compared to conventional SE, Audio-Visual Speech Enhancement (AVSE) additionally incorporates visual information, providing complementary cues that enhance denoising performance \cite{ephrat2018looking, gabbay2017visual, gogate2020cochleanet, michelsanti2021overview}. Prior studies have demonstrated that integrating visual modality significantly improves overall performance \cite{hou2018audio,sadeghi2020audio, kalkhorani2025av, saleem2025viseme}. However, most AVSE systems are trained using conventional objectives such as Scale-Invariant Signal-to-Noise Ratio (SI-SNR) \cite{le2019sdr} and Mean Squared Error (MSE). Although effective for optimization, these objectives do not necessarily align with human subjective perception, creating a gap between training targets and actual listening experience.

Recent Generative Adversarial Network (GAN)-based approaches attempt to incorporate evaluation metrics directly into training. MetricGAN \cite{fu2019metricgan, fu2021metricgan} designs a learnable discriminator to approximate and optimize Perceptual Evaluation of Speech Quality (PESQ) \cite{rix2001perceptual}, while CMGAN \cite{cao2022cmgan} refines both magnitude and phase to improve speech naturalness. Despite the widespread use of metrics such as PESQ, Short-Time Objective Intelligibility (STOI) \cite{taal2010short}, and Scale-Invariant Signal-to-Distortion Ratio (SI-SDR) \cite{le2019sdr}, higher scores do not always correspond to better perceived quality. Improvements in SI-SNR, for example, may still introduce artifacts or unnatural distortions not fully captured by objective measures. The emergence of Large Language Models (LLMs) \cite{brown2020language} provides new opportunities for perceptual assessment. Recent audio LLMs demonstrate strong capabilities in evaluating speech quality \cite{zezario2025study, chen2025audio}. Given a prompt, such models generate natural language descriptions addressing clarity, noise, and distortion \cite{wang2025qualispeech}. These textual evaluations complement traditional metrics and can be converted into numerical scores via sentiment analysis, enabling their integration into training objectives.

In this work, we propose an LLM-based reinforcement learning framework for AVSE, termed LR-AVSE. A pretrained LLM generates perceptually aligned speech quality assessments that serve as reward signals during optimization. Compared with prior RL-based SE approaches—such as using sound quality measures \cite{koizumi2017dnn}, automatic speech recognition performance \cite{shen2019reinforcement}, predicted MOS of NISQA \cite{mittag2021nisqa, kumar2025using}, or direct preference optimization with a neural MOS predictor \cite{li2025aligning}—our method introduces LLM-generated natural language descriptions as rewards, extending perceptual alignment to the audio-visual setting while providing interpretability beyond scalar metrics.

To our knowledge, the proposed LR-AVSE is the first framework to convert LLM-generated descriptive evaluations into reward signals for AVSE optimization. This enables training guided by explicit, human-interpretable explanations. The LLM remains frozen during training to ensure a stable reward criterion. Unlike concurrent work \cite{li2025aligning}, which relies solely on scalar-valued objectives, our approach first produces textual quality descriptions and then converts them into numerical scores, preserving the semantic basis of evaluation. Beyond optimizing scores, this design enhances interpretability: the textual feedback clarifies why a sample is rated higher, such as improved clarity, reduced noise, or diminished distortion. By incorporating natural language quality feedback into SE training, our framework moves beyond black-box numerical optimization toward perceptually grounded and explainable learning.

\section{Methodology}

\begin{figure}[!t]
  \centering
  \includegraphics[width=0.8\linewidth]{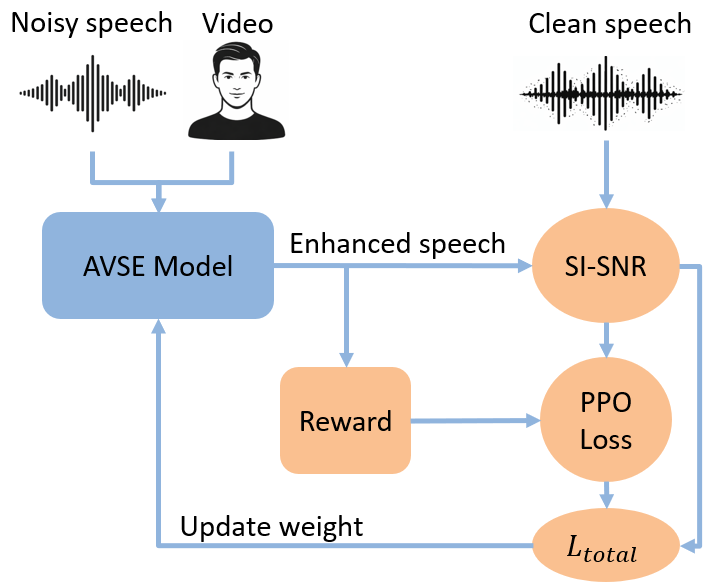}
  \caption{The training procedure of the proposed LR-AVSE framework.} 
  \label{fig:RL}
\end{figure}

This section describes in detail the technical architecture of the proposed LR-AVSE approach. We first introduce the conventional AVSE model as the base architecture, then reformulate the SE problem as a reinforcement learning problem, and describe the RL-based policy optimization strategy and reward model design. The overall system pipeline and architecture are illustrated in Figure~\ref{fig:RL}.

\begin{figure}[!t]
  \centering
  \includegraphics[width=\linewidth]{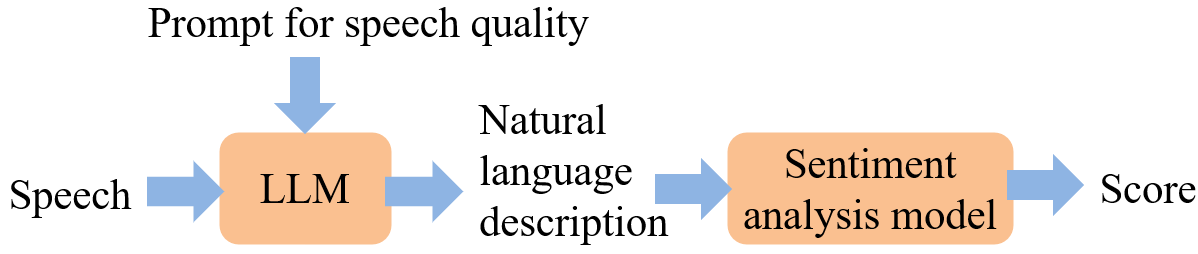}
  \caption{Pipeline of the LLM-based interpretable reward generation.}
  \label{fig:reward}
\end{figure}

\subsection{The AVSE Model}

The AVSE model in our framework is an encoder–separator–decoder architecture adopted in the AVSEC-4 model \cite{avse_challenge_2024}. Let $f_\theta(\cdot)$ denote the neural network with parameters $\theta$, which takes audio and visual inputs and produces the enhanced speech. The noisy speech waveform is denoted as $x \in \mathbb{R}^{C \times T}$, where $C$ denotes the number of channels and $T$ denotes the temporal length. The corresponding visual input is represented as $v \in \mathbb{R}^{1 \times F \times H \times W}$, where $F$ is the number of frames, $H$ and $W$ denote the height and width of each video frame, respectively.

First, the encoder transforms the time-domain signal into a high-dimensional representation:
\[
W = \mathrm{ReLU}(\mathrm{Conv1D}(x)), \quad W \in \mathbb{R}^{B \times N \times K},
\tag{1}
\]
where $N$, $K$, and $B$ are the channel size, the length, and the batch size, respectively.

In the visual branch, visual features are extracted from the video via $\mathrm{VisualFrontend}(\cdot)$:
\[
V = \mathrm{VisualFrontend}(v), \quad V \in \mathbb{R}^{B \times D_v \times T_v},
\tag{2}
\]
where $D_v$ denotes the visual feature dimension, and $T_v$ denotes the temporal length.

The separator adopts a Temporal Convolutional Network (TCN) architecture \cite{luo2019conv}. The visual features are first processed by depthwise temporal convolution and pointwise convolution, then fused with the audio features and fed into the TCN, which ultimately predicts a time-domain mask,
\[
\hat{M} \in \mathbb{R}^{B \times C \times N \times K}.
\]

The decoder applies the predicted mask to the encoder's audio features to reconstruct the corresponding time-domain waveform $\hat{y} \in \mathbb{R}^{C \times T}$:
\[
\hat{y} = \mathrm{Decoder}(W \odot \hat{M}).
\tag{3}
\]

During training, the baseline uses SI-SNR as the optimization objective. Given the clean speech $s$ and the model's estimated speech $\hat{y}$, the loss function is defined as:
\[
\mathcal{L}_{\text{SI-SNR}} = -\mathrm{SI\text{-}SNR}(s, \hat{y}).
\tag{4}
\]

\subsection{SE as a Reinforcement Learning Problem}

We define the supervised fine-tuned model as
$
\pi^{\mathrm{base}}_{\theta} : \mathcal{S} \rightarrow \mathcal{A},
$
where $\mathcal{S}$ denotes the state space, encompassing all possible noisy waveform distributions and thus constituting a continuous space, and $\mathcal{A}$ denotes the action space, comprising all possible mask distributions. In this formulation, the predicted mask $\hat{M}$ is interpreted as the action. Since the original mask output is deterministic, we inject Gaussian noise into the mask to satisfy the stochasticity requirement of reinforcement learning, yielding the RL mask $\hat{M}^{\mathrm{RL}}$:
\[
\hat{M}^{\mathrm{RL}} = f_{\theta}(x, v) + n, \quad n \sim \mathcal{N}(0, \sigma^{2}),
\tag{5}
\]
where $\sigma$ is the standard deviation controlling the degree of stochasticity, which determines the magnitude of noise added to the mask. The RL-optimized policy is denoted as $\pi^{\mathrm{RL}}_{\theta}$, while the original pretrained policy is denoted as $\pi^{\mathrm{base}}_{\theta}$.

\subsection{Reward Model Design}

%
As illustrated in Figures~\ref{fig:RL} and~\ref{fig:reward}, the AVSE model outputs enhanced speech, which is evaluated by the reward model to update the parameters jointly with the Proximal Policy Optimization (PPO) loss~\cite{schulman2017proximal,ouyang2022training} and SI-SNR loss.
We employ a speech-language reward model $r_\phi(\cdot)$, 
which uses an LLM to generate textual descriptions from speech, 
followed by a sentiment analysis model that produces a 1--5 sentiment score.

To stabilize training, we use relative improvement as the reward. Let $\hat{y}^{\mathrm{RL}}$ be the output of $\pi^{\mathrm{RL}}_\theta$ and $\hat{y}^{\mathrm{base}}$ be the output of $\pi^{\mathrm{base}}_\theta$. The reward is computed as:
\[
R = r_\phi(\hat{y}^{\mathrm{RL}}) - r_\phi(\hat{y}^{\mathrm{base}}),
\tag{6}
\]
where
\[
r_\phi(\hat{y}) = \mathrm{Sentiment\_analysis}(\mathrm{LLM}(\hat{y})).
\tag{7}
\]

The natural language descriptions generated by the LLM, such as ``the speech is clear but still has slight background noise'' or ``the denoising effect is good but with a slight sense of distortion'', provide interpretability for the reward, making the training process more than mere score optimization.

To validate the effectiveness of the LLM-based interpretable reward, we also implement a comparative method that uses the predicted MOS score of DNSMOS as the reward. DNSMOS \cite{reddy2021dnsmos} is a pretrained speech quality assessment model that directly predicts MOS scores for speech. In this comparative setting, Equation (7) becomes:
\[
r_\phi(\hat{y}) = \mathrm{DNSMOS}(\hat{y}),
\tag{8}
\]
while the remaining training procedure stays identical. This allows us to quantify the advantage of the LLM-based interpretable reward over a conventional scalar reward.

The overall optimization objective is:
\[
\mathcal{L}(\phi)
= R
- \beta \cdot \mathrm{KL}\Big(
\pi^{\mathrm{RL}}_\phi(\hat{y} \mid x),\,
\pi^{\mathrm{base}}_\phi(\hat{y} \mid x)
\Big),
\tag{9}
\]
where $R$ is the relative reward from Equation (6), and $\beta$ controls KL divergence between the RL and base policies, preventing excessive deviation from the original strategy.

Policy updates are based on PPO \cite{schulman2017proximal}, but we further simplify its structure. Since each episode consists of a single-step decision and the relative reward already embeds baseline information, we replace the conventional advantage function $A_t$ with $\mathcal{L}(\phi)$ from Equation (9), eliminating the need for an additional critic network. The PPO clip loss is defined as:

\begin{equation}
\begin{split}
\mathcal{L}_{\text{clip}}(\phi) &= 
\mathbb{E}_{x \sim \mathcal{D}}
\Bigg[
-\min \Bigg(
\frac{\pi_\phi^{\text{RL}}(\hat{y}\mid x)}{\pi_{\phi^-}^{\text{RL}}(\hat{y}\mid x)} \mathcal{L}(\phi), \\
&\quad\text{clip}\Big(
\frac{\pi_\phi^{\text{RL}}(\hat{y}\mid x)}{\pi_{\phi^-}^{\text{RL}}(\hat{y}\mid x)}, 1-\epsilon, 1+\epsilon
\Big) \mathcal{L}(\phi)
\Bigg)
\Bigg],
\end{split}
\tag{10}
\end{equation}
where $\epsilon$ controls the range of the policy update step, $\phi^-$ denotes the parameters from the previous iteration, and $\mathcal{D}$ is the training dataset used for supervised fine-tuning to obtain the initial pretrained policy $\pi^{\mathrm{base}}_\theta$. 

During the fine-tuning stage, we additionally incorporate the original pretraining loss to stabilize the model. The final total loss function is defined as:
\begin{equation}
\mathcal{L}_{\mathrm{total}} = \mathcal{L}_{\mathrm{clip}} + \gamma \, \mathcal{L}_{\mathrm{SI\text{-}SNR}},
\tag{11}
\end{equation}
where $\gamma$ balances the RL and the SI-SNR SE objectives.

\section{Experimental Setup}

\begin{figure*}[t]
  \centering
  \includegraphics[width=\linewidth]{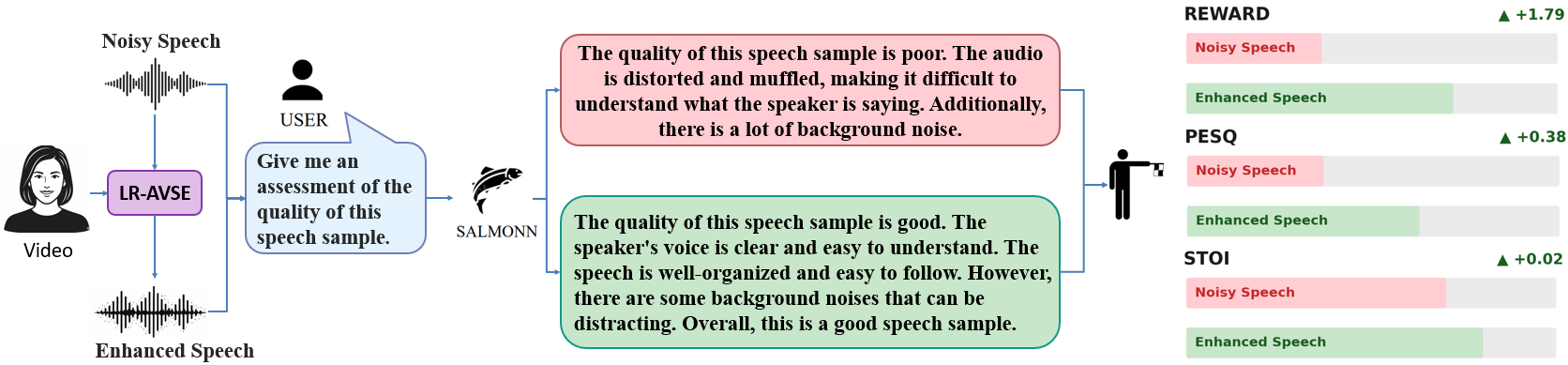}
  \caption{LR-AVSE inference with SALMONN. Rewards from textual descriptions align with PESQ and STOI, demonstrating LR-AVSE’s interpretability.}
  \label{fig:Eval}
\end{figure*}
\subsection{Dataset}

We evaluate the proposed LR-AVSE method on the 4th COG-MHEAR Audio-Visual Speech Enhancement Challenge (AVSEC-4) dataset \cite{avse_challenge_2024}. The training set contains 34,524 scenes, with target speakers drawn from 605 TED/TEDx speakers in the LRS3 \cite{afouras2018lrs3} dataset. Noise sources include 405 competing speakers and 7,346 noise files spanning 15 noise categories, including domestic noises (from CEC1 \cite{cec1}), freesound clips (from DNS v2 \cite{dns_challenge}), and music (from MedleyDB \cite{medleydb}).

The validation set contains 3,365 scenes with 85 target speakers; noise sources include 30 competing speakers and 1,825 noise files. The evaluation set contains 3,180 scenes, and the noises are a subset of those present in the training set. Signal-to-noise ratios range from $-18$\,dB to $+6.55$\,dB. Unlike the simulated room impulse responses used in the training data, the test set employs real recorded room impulse responses captured in three conference rooms at distances of 1 to 2 meters, which poses an additional challenge for model generalization.

All speech signals are downsampled to 16\,kHz with a bit depth of 16. Impulse responses are 6th-order (49-channel) ambisonics signals downsampled to 16\,kHz. In this work, we conduct experiments using binaural signals. Each scene provides silent video, target speech, and their mixed audio.

\subsection{Training}

The training procedure consists of two stages. In the pretraining stage, we initialize the model with the pretrained weights provided by the AVSEC-4 organizers. In the RL fine-tuning stage, the hyperparameters are set as follows: $\sigma = 0.05$, PPO clip range $\epsilon = 0.1$, $\beta = 0.0001$, SI-SNR loss weight $\gamma = 1.0$, and learning rate is set to $ 0.001$.

For the LLM, we adopt SALMONN \cite{tang2023salmonn}, which has been fine-tuned for speech-related tasks, particularly speech quality understanding, and is therefore capable of generating semantically meaningful quality-related descriptions from speech. During training, we use one of the prompts officially recommended by SALMONN: ``Give me an assessment of the quality of this speech sample,'' to guide the model in producing natural language evaluations of speech quality.

The generated textual descriptions are then fed into a sentiment analysis model, specifically BERT \cite{nlptown_bert_sentiment, devlin2019bert}, which converts them into quality scores on a scale of 1 to 5, serving as the final reward signal.

\subsection{Baselines}

We compare our method against two baselines. 
All models are built upon the same pretrained AVSE backbone, which adopts an encoder–separator–decoder architecture and is initially trained with supervised learning using SI-SNR as the objective function. The pretrained model and weights are officially 
provided by the AVSEC-4 organizers.
Reinforcement learning fine-tuning is then applied on top of this pretrained model following the reinforcement learning from human feedback (RLHF) paradigm, where a reward model provides feedback to guide policy optimization via PPO.

\begin{itemize}
\item
\textbf{Pretrained Baseline}: This model uses the AVSEC-4 provided pretrained weights and has been trained solely with supervised learning using SI-SNR as the optimization objective, without any RL fine-tuning. It represents the conventional supervised AVSE approach.
\item
\textbf{RL-DNSMOS}: This baseline follows the same RL fine-tuning pipeline as our method, but replaces the SALMONN+BERT reward model with DNSMOS. DNSMOS is a pretrained speech quality assessment model that directly outputs MOS predictions on a scale of 1 to 5. This comparison allows us to quantify the advantage of the LLM-based interpretable reward over a conventional scalar reward. For fair comparison, RL-DNSMOS uses the same PPO hyperparameters ($\sigma = 0.05$, $\epsilon = 0.1$, $\beta = 0.0001$,  $\gamma = 1.0$) and training procedure as our proposed method.

\end{itemize}

\section{Results}

\subsection{Objective Quality Metrics}

As shown in Table~\ref{tab:test_objective_results}, evaluated on the test set, LR-AVSE achieves a PESQ of 1.25, outperforming the Pretrained Baseline at 1.20 and RL-DNSMOS at 1.24.
Regarding the NISQA-predicted MOS, LR-AVSE reaches 1.29, significantly outperforming the Noisy input at 0.97 and the Pretrained Baseline at 0.99. On the VQscore \cite{fu2024self} and SpeechBERTScore (S-BERT) \cite{shi2025versa, shi2024versaversatileevaluationtoolkit, saeki2024speechbertscore} metrics, LR-AVSE obtains scores of 0.62 and 0.57, respectively, achieving the best overall performance among all compared methods. A comparison with RL-DNSMOS further highlights the advantage of the LLM-based interpretable reward: although both methods share the same RL fine-tuning framework, the LLM-based interpretable reward outperforms the DNSMOS-based reward on PESQ, STOI and neural speech quality assessment scores, indicating that natural language descriptions provide richer quality information that helps the model learn more nuanced enhancement strategies.

\begin{table}[th]
  \caption{Objective results on the AVSEC-4 test set. For PESQ evaluation, we used the wideband (WB) mode.}
  \label{tab:test_objective_results}
  \footnotesize
  \centering
  \setlength{\tabcolsep}{4pt}
  \begin{tabular}{l c c c c c}
    \toprule
    \textbf{Method} & \textbf{PESQ} & \textbf{STOI} & \textbf{NISQA} & \textbf{VQscore} & \textbf{S-BERT} \\
    \midrule
    Noisy     & 1.08 & 0.55 & 0.97 & 0.58 & 0.55 \\
    Baseline  & 1.20 & 0.48 & 0.99 & 0.61 & 0.54 \\
    RL-DNSMOS & 1.24 & 0.57 & 1.15 & 0.62 & 0.56 \\
    LR-AVSE   & \textbf{1.25} & \textbf{0.58} & \textbf{1.29} & \textbf{0.62} & \textbf{0.57} \\
    \bottomrule
  \end{tabular}
\end{table}

\subsection{Subjective Evaluation}

\begin{figure}[!t]
  \centering
  \includegraphics[width=0.85\linewidth]{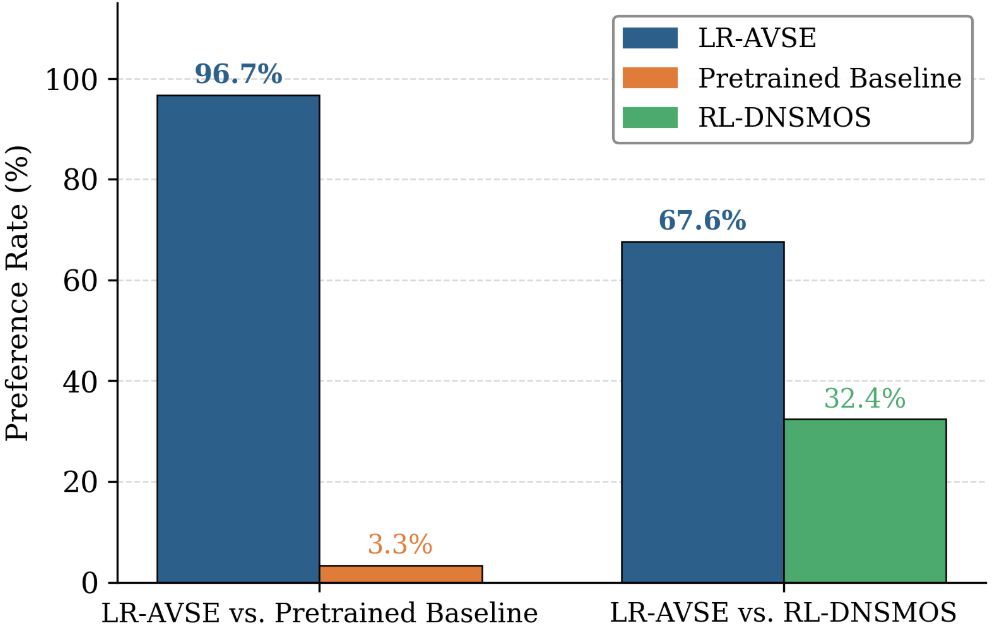}
  \caption{A/B preference test results on the AVSEC-4 test set.}
  \label{fig:ab_preference}
\end{figure}

To subjectively evaluate speech quality, we conducted an A/B preference test. A total of 21 participants were recruited for the experiment. Two comparison conditions were designed: LR-AVSE vs. Pretrained Baseline and LR-AVSE vs. RL-DNSMOS. Each comparison included 10 utterances. As shown in Fig.~\ref{fig:ab_preference}, LR-AVSE achieved a 96.7\% preference rate over the Pretrained Baseline. When compared with RL-DNSMOS, LR-AVSE still obtained a 67.6\% preference rate.

Based on the subjective evaluation results, the effectiveness of the proposed method is further validated. Through the LLM-based interpretable reward mechanism, the framework not only improves speech quality performance but also provides a more interpretable foundation for the model optimization process.

\subsection{Interpretable Reward Analysis}

Figure~\ref{fig:Eval} presents the natural language descriptions generated by SALMONN for noisy speech and enhanced speech. For the noisy speech, SALMONN described it as ``distorted and muffled,'' ``difficult to understand,'' and containing ``a lot of background noise.'' After enhancement using our proposed method, the speech is described as ``clear and easy to understand'' and ``well-organized.'' Although ``some background noises'' are still mentioned, it is overall characterized as a ``good speech sample.'' The corresponding reward increases by $\Delta$+1.79, while PESQ improves by $\Delta$+0.38 and STOI by $\Delta$+0.02.

These results show a consistent improvement trend across the three metrics---Reward, PESQ, and STOI---thereby validating the core hypothesis of this study: the reward derived from LLM-generated natural language descriptions exhibits a positive correlation with conventional objective metrics. Compared to traditional approaches that rely on a single numerical score to assess speech quality, the descriptions provided by SALMONN offer more insights into the aspects of improvement, such as enhanced intelligibility, reduced distortion, and decreased background noise. Such feedback allows for a clearer understanding of how the model improves speech quality, making the evaluation process more intuitive and interpretable.

\section{Discussion}

Our LLM-based reward demonstrates clear advantages in interpretability; however, the current choice of LLM remains relatively limited. When generating natural language descriptions, the produced sentences tend to follow fixed patterns—for example, repeating phrases such as ``The quality of this speech sample is poor/good'' or ``The audio is distorted and muffled.'' These repetitive descriptions impose a certain constraint on our proposed LLM-based reward, potentially making it difficult to capture subtle differences in speech quality.

Future research directions may include: (1) adopting more advanced LLMs as the reward model, as models with stronger capabilities can generate richer vocabulary and more nuanced natural language descriptions of speech quality differences; (2) more careful prompt engineering during training—for instance, using more detailed prompts such as ``Please evaluate the speech in terms of clarity, noise level, timbral naturalness, and loudness stability'' to guide the model toward producing more structured and fine-grained descriptions.


\section{Conclusion}

This paper proposes LR-AVSE, which, to the best of our knowledge, is the first framework that leverages reinforcement learning from LLM feedback for AVSE. The proposed LLM-based interpretable reward achieves notable improvements over the supervised-trained baseline and DNSMOS-based RL AVSE systems across PESQ, STOI, neural speech quality assessment metrics, and subjective listening tests. Furthermore, unlike conventional black-box training paradigms that rely solely on single numerical values, the proposed method provides interpretability. Even with current limitations of LLMs, the natural language form of the LLM-based interpretable reward can effectively guide the model toward continuous improvement. Future work will explore the use of more advanced LLMs, improved prompt design strategies, and the extension of the proposed framework to a broader range of speech processing tasks.

\section{Generative AI Use Disclosure}
Generative AI was used only for editing and polishing this manuscript.


\bibliographystyle{IEEEtran}
\bibliography{mybib}

\end{document}